\documentclass[twocolumn,aps,tightenlines,floatfix,showpacs]{revtex4}
\usepackage{bm}% bold math
\usepackage{graphics}
\usepackage{graphicx}
\usepackage{epsfig}
\usepackage{longtable}
\newcommand{\btab}{\begin{tabular}}
\newcommand{\etab}{\end{tabular}}
\usepackage{multirow,amsmath,booktabs}
\begin{document}

\title{Renormalized interactions with a realistic single particle basis}

\author{Angelo Signoracci$^{1}$}
\author{B. Alex Brown$^{1}$}
\author{Morten Hjorth-Jensen$^{2}$}

\affiliation{$^{1}$Department of Physics and Astronomy
and National Superconducting
Cyclotron Laboratory,
Michigan State University,
East Lansing, Michigan 48824-1321, USA}

\affiliation{$^{2}$Department of Physics and Center of Mathematics for 
Applications,
 University of Oslo, N-0316 Oslo, Norway}

\begin{abstract}
Neutron-rich isotopes in the $  sdpf  $ space with $  Z \leq 14  $ 
require modifications to derived effective
interactions to agree with experimental data away from stability.  A
quantitative justification is given for these modifications due to the 
weakly bound nature of model space
orbits via a procedure using realistic radial wavefunctions and 
realistic $  NN  $ interactions.  The long tail of
the radial wavefunction for loosely bound single particle orbits causes 
a reduction in the
size of matrix elements involving those orbits, most notably for 
pairing matrix elements,
resulting in a more condensed level spacing in shell model 
calculations.  Example
calculations are shown for $^{36}$Si and $^{38}$Si.
\end{abstract}


\date{\today}
\maketitle

\section{Introduction}

New facilities for rare isotope beams will push the experimental 
capabilities
of nuclear physics with radioactive beams to more unstable, 
shorter-lived nuclei.
Properties of these nuclei exhibiting different behavior than stable 
nuclei,
like the evolution of shell structure, are of significant interest for 
the next decades
of research.  A new theoretical technique and its behavior for stable 
and exotic nuclei has been
studied to examine the importance of refining theoretical approaches 
for the production
of model space interactions for unstable nuclei.

Much research has been done using renormalization methods to convert a 
realistic interaction
fit to nucleon-nucleon ($  NN  $) scattering data into an interaction 
in the nuclear medium.  The goal is to
renormalize the interaction to valence orbits outside of a stable,
semi-magic or doubly magic nucleus treated as a vacuum in further 
calculations.
A typical example would use $^{16}$O as the core and renormalize the $  
NN  $ interaction into the
$  sd  $ model space.  For such an application, the harmonic oscillator 
basis of the form
$  \Psi _{nlm_{l}}  $($  \vec{r}  $) = $  R^{HO}_{nl}  $(r) $  
Y_{lm_{l}}  $($\theta$,$\phi$) is generally used.  Additionally, all the
valence orbits are bound in the harmonic oscillator basis.
For more exotic closed-subshell nuclei, loosely
bound orbits often play a role.  The harmonic oscillator basis is less 
applicable
further from stability.  Loosely bound orbits
particularly deviate from the oscillator basis, as they exhibit a 
``long-tail" behavior with a
larger spread in the radial wavefunctions.  However, few calculations 
have
been done with a realistic radial basis for unstable nuclei with 
renormalized $  NN  $ interactions.

Experimental interest in neutron-rich silicon isotopes and the failure 
of some shell
model Hamiltonians to reproduce data in the region have led to 
modifications in the SDPF-NR
interaction \cite{now}, which had been the standard for shell model 
calculations in the $  sdpf  $ model
space.  The new SDPF-U interaction has different neutron-neutron 
pairing matrix elements
for $  Z \geq 15  $ and $  Z \leq 14  $ to account for the behavior of 
$  pf  $ neutron orbits relative to
the number of valence protons.  The $  Z \leq 14  $ version of the 
interaction treats neutron-rich
unstable nuclei that exhibit different shell behavior than the less 
exotic nuclei in the
$  Z \geq 15  $ nuclei.  The interest in silicon isotopes and the 
nature of the SDPF-U interaction
make $^{34}$Si a suitable choice for the renormalization procedure with 
a realistic basis.  A similar effect
occurs for the neutron-rich carbon isotopes around the $  N=14  $ 
closed subshell, requiring a 25\%
reduction in the neutron-neutron two-body matrix elements from the 
effective interactions derived
for the oxygen isotopes \cite{stan}.

\section{Renormalization Procedure}

We begin with the realistic charge-dependent $  NN  $ interaction 
N$^{3}$LO derived
at fourth order of chiral perturbation theory with a 500 MeV cutoff
and fit to experimental $  NN  $ scattering data \cite{n3lo}.  The 
N$^{3}$LO interaction is
renormalized using a similarity transformation in momentum space
with a sharp cutoff of $  \Lambda  $ = 2.2 fm$^{-1}$ to obtain the 
relevant low momentum interaction
\cite{vlowk}.  We will refer to this technique as a $  v_{low k}  $ 
renormalization.  Skyrme Hartree-Fock
calculations are performed with the Skxtb interaction \cite{skxtb} for 
a chosen closed sub-shell target
nucleus to determine the binding energy, single particle radial 
wavefunctions, and single particle
energy spectra for neutrons and protons of the target nucleus.  The low
momentum interaction is then renormalized into a model space of 
interest using Rayleigh-Schr\"odinger
perturbation theory \cite{morten} to second order including excitations 
up to 6$  \hbar  \omega$, summing
over folded diagrams to infinite order.  We will compare three options 
for the renormalization
to produce an effective interaction:
harmonic oscillator single particle energies and wavefunctions (HO),
Skyrme Hartree-Fock single particle energies and wavefunctions (SHF), 
and
Skyrme Hartree-Fock single particle energies and harmonic oscillator 
single particle wavefunctions (CP).

The CP basis and HO basis give identical results to first order in 
perturbation theory since they use
identical wavefunctions.  The energies, which are different in the two 
procedures, come into higher
order diagrams via energy denominators, as discussed in \cite{morten}.  
Therefore, the
last option is a core-polarization basis (CP basis) since the 
core-polarization diagrams
are affected to leading order even though the result at first order is 
unchanged.

Skyrme Hartree-Fock radial wavefunctions, once solved, are implemented 
in the renormalization
by using an expansion in terms of the harmonic oscillator basis via:
$$
\psi ^{SHF}_{nlj}(\vec{r}) = \sum_{{}n} a_{n} R^{HO}  _{nl}  (r) 
[Y_{l}(\theta ,\phi ) \otimes \chi _{s}]_{j},  \eqno({1})
$$
where $  a_{n}^{2}  $ gives the percentage of a specific harmonic 
oscillator wavefunction component in the
Skyrme Hartree-Fock wavefunction.  The Skyrme Hartree-Fock 
wavefunctions and single particle energies can only be
determined for bound states.  For unbound orbits, the harmonic 
oscillator basis remains in use, but the
Gram-Schmidt process is used to ensure orthonormality of the single 
particle wavefunctions.  The
effective interaction, consisting of the derived two-body matrix 
elements and the Skyrme Hartree-Fock
single particle energies, can then be used in a shell model program 
directly.

\section{Application to sdpf model space}

\begin{table}
\begin{center}
\caption{Single-particle energies for $^{34}$Si and
$^{40}$Ca using the Skxtb interaction.
Values in bold are in the model space.}
\begin{tabular}{|r||r|r|r|r|}
\hline n$  l_{j}  $ & $^{34}$Si & $^{34}$Si & $^{40}$Ca & $^{40}$Ca\\
 & proton & neutron & proton & neutron \\
\hline
\hline
 $  0s_{1/2}  $ & -37.73 & -32.79 & -30.49 & -38.18 \\
 $  0p_{3/2}  $ & -27.60 & -23.10 & -22.14 & -29.70 \\
 $  0p_{1/2}  $ & -22.39 & -21.74 & -19.03 & -26.67 \\
 $  0d_{5/2}  $ & {\bf -17.29} & -13.07 & {\bf -12.79} & -20.20 \\
 $  0d_{3/2}  $ & {\bf -9.08} &  -9.03 &  {\bf -7.23} & -14.65 \\
 $  1s_{1/2}  $ & {\bf -13.49} & -10.04 & {\bf -8.31} & -15.75 \\
 $  0f_{7/2}  $ &  -5.97 & {\bf -2.62} &  -2.68 & {\bf -9.89} \\
 $  0f_{5/2}  $ &   3.70 &  {\bf 3.33} &   4.81 & {\bf -2.43} \\
 $  1p_{3/2}  $ &  -1.06 &  {\bf -0.40} &   1.44 & {\bf -5.48} \\
 $  1p_{1/2}  $ &   1.49 &  {\bf -0.27} &   3.27 & {\bf -3.66} \\
 $  0g_{9/2}  $ &   6.39 &   9.22 &   8.63 &   1.15 \\
 $  0g_{7/2}  $ &  18.26 &  18.23 &  18.76 &  10.28 \\
\hline
\end{tabular}
\end{center}
\end{table}

\begin{table}
\begin{center}
\caption{Single-particle energies for $^{34}$Si and
$^{40}$Ca in the harmonic oscillator basis.  The energy shift
is chosen so that the valence energy is identical in both bases.
Values in bold are in the model space.}
\begin{tabular}{|r||r|r|r|r|}
\hline n$  l_{j}  $ & $^{34}$Si & $^{34}$Si & $^{40}$Ca & $^{40}$Ca\\
 & proton & neutron & proton & neutron \\
\hline
\hline
 $  0s_{1/2}  $ & -36.93 & -34.59 & -32.22 & -39.21 \\
 $  0p_{3/2}  $ & -25.42 & -23.09 & -21.20 & -28.19 \\
 $  0p_{1/2}  $ & -25.42 & -23.09 & -21.20 & -28.19 \\
 $  0d_{5/2}  $ & {\bf -13.91} & -11.58 & {\bf -10.18} & -17.17 \\
 $  0d_{3/2}  $ & {\bf -13.91} & -11.58 & {\bf -10.18} & -17.17 \\
 $  1s_{1/2}  $ & {\bf -13.91} & -11.58 & {\bf -10.18} & -17.17 \\
 $  0f_{7/2}  $ &  -2.40 &  {\bf -0.07} &   0.84 &  {\bf -6.15} \\
 $  0f_{5/2}  $ &  -2.40 &  {\bf -0.07} &   0.84 &  {\bf -6.15} \\
 $  1p_{3/2}  $ &  -2.40 &  {\bf -0.07} &   0.84 &  {\bf -6.15} \\
 $  1p_{1/2}  $ &  -2.40 &  {\bf -0.07} &   0.84 &  {\bf -6.15} \\
 $  0g_{9/2}  $ &   9.11 &  11.44 &  11.86 &   4.87 \\
 $  0g_{7/2}  $ &   9.11 &  11.44 &  11.86 &   4.87 \\
\hline
\end{tabular}
\end{center}
\end{table}

Neutron-rich silicon isotopes present an interesting
application of the procedure outlined in the last section.  A deeper 
understanding of
the need for multiple interactions in the $  sdpf  $ model space, as 
seen by the form of
SDPF-U, can be gained by performing the renormalization for the same 
model space in
multiple ways.  The model space chosen is the $  sd  $ proton orbits 
and $  pf  $ neutron
orbits.  The renormalization procedure is done using all three
options for two different target nuclei, producing a total of six 
interactions.
The two target nuclei chosen are the stable $^{40}$Ca doubly magic 
nucleus, and the
neutron-rich $^{34}$Si semi-magic nucleus.  Single particle energies of 
the SHF basis,
using the Skxtb interaction,  are presented in Table I for both target 
nuclei.
For an SHF state that is unbound, the radial wavefunction is 
approximated by a state
bound by 200 keV that is obtained by multiplying the SHF central 
potential by a factor
larger than unity. The energy of the unbound state is estimated by 
taking the
expectation value of this bound state wavefunction in the original SHF 
potential.

In the SHF basis, the calculation of single particle energies shows 
that the
proton orbits are shifted down in energy for $^{34}$Si relative to 
$^{40}$Ca, while the
neutron orbits are shifted up.  For the valence neutrons, this shift 
results in a switch
from four orbits for $^{40}$Ca bound by 5.4 Mev on average to four 
orbits for $^{34}$Si
centered at 0.0 MeV.  This change, specifically the loosely bound 
energies of the
$  p_{3/2}  $ and $  p_{1/2}  $, has a significant effect on the 
wavefunctions, which will be
discussed in more detail later.
For comparison,
the single particle energies used in the HO basis are
given in Table II.  The Blomqvist-Molinari formula \cite{bm}
$  \hbar \omega = (45 A^{-1/3} - 25 A^{-2/3})  $ MeV
gives 11.508 MeV for $  A =   $34 and 11.021 MeV for $  A =  $ 40.  The 
absolute value
of the harmonic oscillator basis is irrelevant, as only energy 
differences come into the
diagrams in Rayleigh-Schr\"odinger perturbation theory.  For a better 
comparison to the
SHF basis, the absolute value is chosen separately for protons and 
neutrons such that
$  \sum\limits_{1}^{n_{val}} (2J+1) \epsilon_{\alpha}  $ is identical 
in the HO and SHF bases,
where $  n_{val}  $, the number of valence orbits, is three for protons 
and four for
neutrons and $  \epsilon_{\alpha}  $ is the energy of the single 
particle orbit given by the
$  \alpha = n,l,j  $ quantum numbers.  In order to avoid divergences 
from the calculation of
energy denominators, all model space orbits are set to the same valence 
energy such that the
starting energy \cite{morten} of each diagram is constant.

\begin{figure}
\scalebox{0.5}{\includegraphics{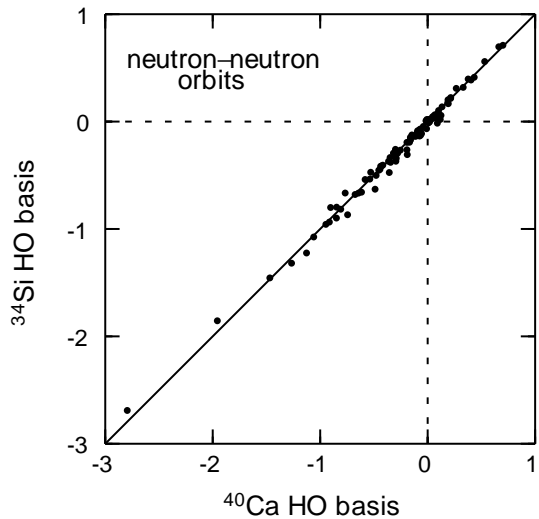}}
\caption{Comparison of $  pf  $ neutron-neutron matrix elements (in 
MeV) for the
renormalization procedure in the HO basis for the two target
nuclei.  The solid line $  y=x  $ denotes where the matrix elements 
would be identical. }
\label{(1)}
\end{figure}

\begin{figure}
\scalebox{0.5}{\includegraphics{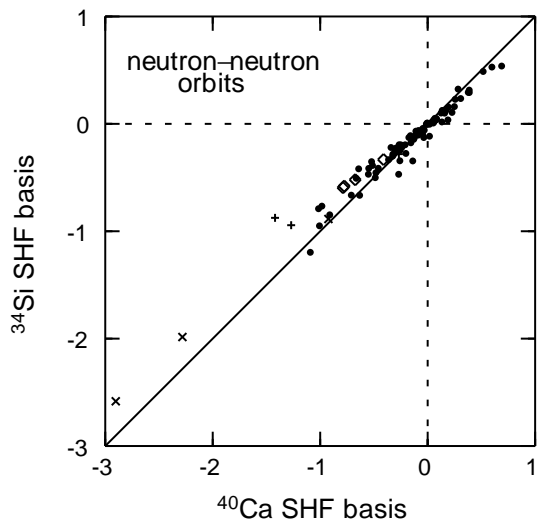}}
\caption{Comparison of $  pf  $ neutron-neutron matrix elements (in 
MeV) for the
renormalization procedure in the SHF basis for the two target nuclei.  
The solid line $  y=x  $ denotes
where the matrix elements would be identical. Black dots correspond to 
matrix elements with
$  J > 0  $, while the $  J = 0  $ matrix elements are split into three 
groups:
$  ff-ff  $ (crosses), $  ff-pp  $ (diamonds), and $  pp-pp  $ (plus 
signs).}
\label{(2)}
\end{figure}

\begin{figure}
\scalebox{0.5}{\includegraphics{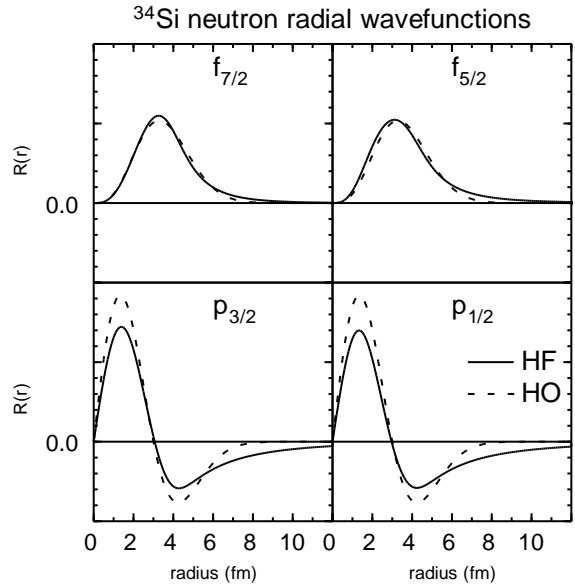}}
\caption{Comparison of the single particle radial wavefunctions for 
$^{34}$Si in the HO and
SHF bases.}
\label{(3)}
\end{figure}

Fig.\ 1 shows a comparison of the $  pf  $ matrix elements in MeV for 
both target nuclei
with the HO basis used in the renormalization procedure.  The
values deviate slightly from the line of equality but agree well with 
each other.
Therefore, the choice of target nucleus, whether $^{34}$Si or 
$^{40}$Ca, has little effect on the matrix
elements in the HO basis.  However, when we look at
Fig.\ 2, where the comparison is for both target nuclei in
the SHF basis, we see a reduction in the strength of the interaction 
for $^{34}$Si.
This reduction with $^{34}$Si as the target
nucleus is due to the weakly bound nature of the $  pf  $ neutron 
orbits.

In the SHF basis, the $  f_{7/2}  $ orbit is bound by 2.6 MeV, and its 
radial wavefunction agrees
well with the harmonic oscillator wavefunction as seen in Fig.\ 3.  The 
Skyrme wavefunction is
expanded in the harmonic oscillator basis up to $  n=n_{max}  $ and the 
$  a_{n}  $ coefficients
are renormalized to ensure that $  \sum\limits_{n=0}^{n_{max}} 
a_{n}^{2} = 1  $.
For our renormalization procedure, orbits up to $  (2n+l)=9  $ are 
included,
which gives $  n_{max}=3  $ for the $  f_{7/2}  $ and $  f_{5/2}  $ 
orbits and $  n_{max}=4  $
for the $  p_{3/2}  $ and $  p_{1/2}  $ orbits.  This includes over 
99\% of the strength for the $  f  $
orbits, but only 93\% and 92\% for the $  p_{3/2}  $ and $  p_{1/2}  $ 
orbits respectively.  A first order
calculation can be done to $  n_{max}=6  $ for all orbits, which gives 
100\%, 98\%, and 97\% for the
$  f  $, $  p_{3/2}  $, and $  p_{1/2}  $ expansions respectively.

With this procedure, 99\% of the $  f_{7/2}  $ orbit is represented by 
the $  R^{HO}_{03}  $ wavefunction.
The $  p_{3/2}  $ and $  p_{1/2}  $ orbits are only bound by 400 and 
269 keV, respectively.
The expected harmonic oscillator component $  R^{HO}_{11}  $ only makes 
up 80\% and 78\% of the
respective radial wavefunctions.  Higher $  n  $ orbits which extend 
farther away from the center
of the nucleus contribute the remaining strength.  The $  f_{5/2}  $ 
orbit is unbound by three MeV, but the
solution for the Skyrme radial wavefunction is determined by assuming 
that the orbit is bound by
200 keV.  With this method, 97\% of the realistic radial wavefunction 
is given by the $  R^{HO}_{03}  $
wavefunction.  Single particle radial wavefunctions of valence space 
neutron orbits are shown in Fig.\ 3
in both the HO and SHF basis.
The long tail behavior of the loosely bound $  p  $ orbits can be seen 
in the SHF basis,
as the wavefunctions have significant strength beyond 8 fm unlike the 
oscillator wavefunctions.

The $  J=0  $ matrix elements in Fig.\ 2 deviate more from the line of 
equality, i.e.
the pairing matrix elements are reduced
for $^{34}$Si when the N$^{3}$LO interaction is renormalized in the SHF 
basis.
The SDPF-U interaction has different neutron-neutron pairing matrix 
elements for $  Z \geq 15  $
and for $  Z \leq 14  $ to account for 2p-2h excitations of the core 
correctly, depending on whether
$^{34}$Si or $^{40}$Ca should be considered the core \cite{now}.  The 
SDPF-U neutron-neutron pairing matrix elements are
reduced by 300 keV for $  Z \leq 14  $ in order to produce results in 
better agreement with experimental data.
The pairing matrix elements in Fig.\ 2 are reduced for the $^{34}$Si 
target by 213 keV on average,
relative to the case with $^{40}$Ca as the target.  While the 
connection here to the $  Z  $-dependence in
SDPF-U is only suggestive, the change in target mimics the change in 
core for calculations in the $  sdpf  $ region, cited
by Nowacki and Poves as the cause of their 300 keV reduction \cite{now}.
The reduction of 213 keV is due solely to the change in occupation of
the $  d_{5/2}  $ proton orbit, which can affect the single particle 
energies and radial wavefunctions, as well as
the available diagrams in the core polarization.  We find that the 
change in single particle radial wavefunctions plays
the most significant role, but are also able to analyze the effect of 
the core polarization.

\begin{table}
\begin{center}
\caption{Core polarization and total matrix elements in MeV of the form
$  \langle a a \hspace{.04cm} \vline \hspace{0.03cm} V \hspace{0.03cm} 
\vline \hspace{0.05cm} b b \rangle _{J=0}  $
for different renormalization procedures.}
\begin{tabular}{|c|c|c||c|c|c||c|c|c|}
\hline
\multicolumn{3}{|c||}{} & \multicolumn{3}{|c||}{$^{34}$Si} & 
\multicolumn{3}{|c|}{$^{40}$Ca} \\
\hline  a & b &  & HO & CP & SHF & HO & CP & SHF \\
\hline
\hline
 f$_{7/2}$ & f$_{7/2}$ & core pol. & -0.449 & -0.377 & -0.529 & -0.637 
& -0.649 & -0.931 \\
  &  & total & -1.855 & -1.869 & -1.985 & -1.957 & -1.982 & -2.282 \\
\hline
 p$_{3/2}$ & p$_{3/2}$ & core pol. & -0.037 &  0.001 &  0.010 & -0.021 
& -0.005 & -0.015 \\
  &  & total & -1.319 & -1.313 & -0.944 & -1.267 & -1.252 & -1.270 \\
\hline
 p$_{3/2}$ & p$_{1/2}$ & core pol. & 0.068 & 0.082 & 0.047 & -0.038 & 
-0.069 & -0.087 \\
 &   & total & -1.456 & -1.462 & -0.875 & -1.469 & -1.488 & -1.420 \\
\hline
\end{tabular}
\end{center}
\end{table}

Table III isolates a few matrix elements and
compares the total matrix elements and the component due to core 
polarization for
both target nuclei in all three bases.  The reduction for total matrix 
elements involving
$  p  $ orbits is dramatic ($  \approx  $ 30\%) and is primarily due to 
the extension of
wavefunction strength to large distances.  Kuo et al. \cite{refa} noted 
a reduction of core polarization
in the harmonic oscillator basis and used different oscillator 
parameters to account for the core nucleons
and valence nucleons separately in halo nuclei.
While $^{34}$Si is not a halo nucleus, the loosely bound $  p  $ orbits 
behave in much the same way as the
valence nucleons in a halo nucleus.  The reduction in core
polarization is seen going from the $^{40}$Ca target to the $^{34}$Si 
target in any basis in Table III,
although the size of the polarization is reduced for nucleons far from 
the core.  As noted in \cite{refa},
the core interacts less with nucleons far away, so the excitations of 
the core are reduced.  The
core polarization for matrix elements solely involving $  p  $ orbits 
is under 100 keV.  We observe that
the core polarization can be reduced significantly without the total 
matrix element changing in the
same proportion.  For instance, the
$  \langle f_{7/2} f_{7/2} \hspace{0.04cm} \vline V \hspace{0.04cm} 
\vline f_{7/2} f_{7/2} \hspace{0.02cm} \rangle  $
matrix element is only reduced by 5\% from $^{40}$Ca to $^{34}$Si in 
the HO basis even though the core
polarization is reduced by 30\%.  In the SHF basis, which takes into 
account the realistic
wavefunction, the total matrix element is reduced by 13\% even though 
the core polarization is
reduced by 43\%.  We would prefer to compare matrix elements involving 
the $  p_{3/2}  $ or $  p_{1/2}  $
orbits, but the core polarization becomes very small for loosely bound 
orbits, skewing percentage
comparisons.  Ogawa et al. \cite{refb} produce results which seem to be 
consistent with ours, identifying a
10\%-30\% reduction in nuclear interaction matrix elements involving 
loosely bound orbits using a realistic Woods-Saxon basis.
However, they were limited to comparisons of ratios of matrix elements 
and did not
include core polarization.  We show that
core polarization suppression and reduction due to spread of the 
wavefunctions are both important effects
which should be included, but do not tell the entire story.  The $  
f_{7/2}  $ wavefunction is very similar in the
HO and SHF bases, as seen in Fig.\ 3, and yet the
$  \langle f_{7/2} f_{7/2} \hspace{0.04cm} \vline V \hspace{0.04cm} 
\vline f_{7/2} f_{7/2} \hspace{0.02cm} \rangle  $
matrix element does not follow the same trend as the core polarization 
component in Table III.  Other diagrams
which are included at second order are relevant, and the full treatment 
of the renormalization in a realistic basis, as
developed here, is necessary for accurate results.  Our improvements 
enable us to perform calculations for
neutron-rich silicon isotopes directly.

\section{Calculations for $^{36}$Si and $^{38}$Si}

The effect of the different interactions on nuclear structure 
calculations has been studied
as neutrons are added to $^{34}$Si.  In order to obtain a consistent
starting point, the proton-proton and proton-neutron matrix elements of 
SDPF-U have been used, with proton single
particle energies (SPEs) chosen to reproduce those obtained by SDPF-U.  
Because SDPF-U does not
reproduce the binding energy of $^{35}$Si, the SDPF-U neutron SPEs have 
been increased by 660 keV.
The six interactions use neutron SPEs that reproduce the values of this 
modified
SDPF-U interaction.

\begin{figure}
\scalebox{0.5}{\includegraphics{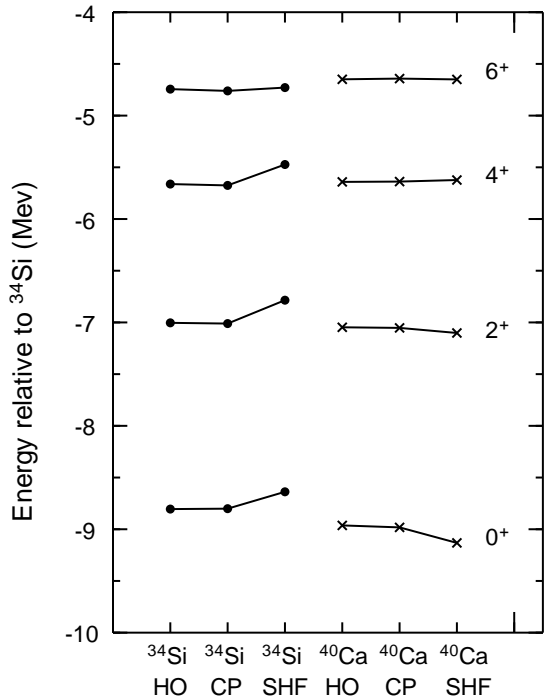}}
\caption{Calculations for the lowest energy states for $  J=0,2,4,6  $ 
in $^{36}$Si
relative to $^{34}$Si from the renormalization procedure for $^{34}$Si 
and $^{40}$Ca,
in the HO, CP, and SHF bases for both target nuclei.
Crosses are used for calculations with $^{40}$Ca as the target nucleus.}
\label{(4)}
\end{figure}

The only difference in the six interactions used in the calculations 
are the neutron-neutron
matrix elements.  Calculations have been done in the model space 
discussed in the last section
with the shell model code NuShellx \cite{msunu}.
Fig.\ 4 shows the lowest $  J=0,2,4,6  $ states in $^{36}$Si, relative 
to $^{34}$Si.  The HO
basis and the CP basis for the same target nucleus deviate by no more 
than 20 keV.
However, the SHF basis noticeably shifts the states, with the largest 
effect
being 170 keV more binding in the ground state with $^{40}$Ca as the 
target nucleus.
The binding energy of $^{36}$Si changes by nearly 500 keV depending on 
which
renormalization procedure is used.  Furthermore, the level schemes for 
$^{36}$Si are
more spread out for the crosses where $^{40}$Ca is chosen as the target 
nucleus.

\begin{figure}
\scalebox{0.5}{\includegraphics{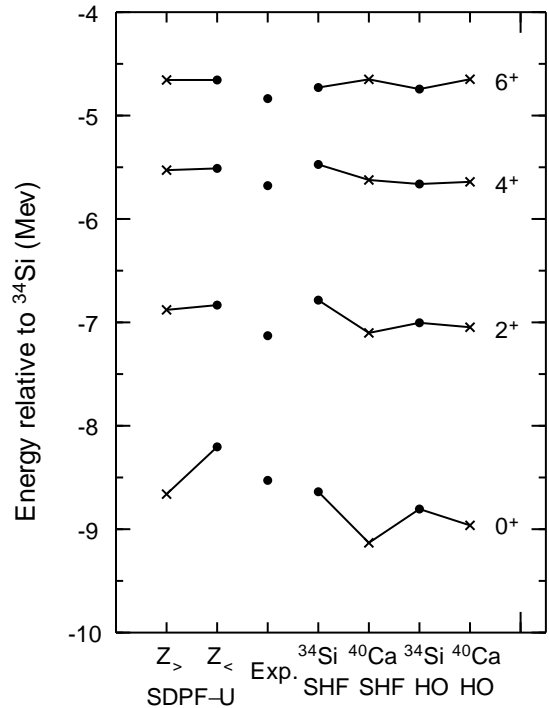}}
\caption{Calculations for the lowest energy states for $  J=0,2,4,6  $ 
in $^{36}$Si
relative to $^{34}$Si using neutron-neutron matrix
elements from SDPF-U and the renormalization procedure for both 
$^{34}$Si and $^{40}$Ca
as target nuclei, using the SHF and HO bases.
Experimental data is shown for comparison, with a new mass from 
\cite{mittig}.\
Crosses are used for calculations with $^{40}$Ca as the target nucleus. 
}
\label{(5)}
\end{figure}

\begin{figure}
\scalebox{0.5}{\includegraphics{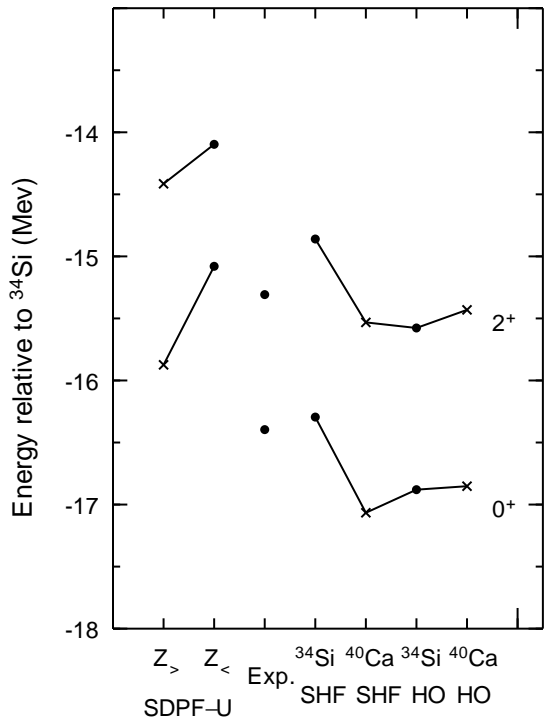}}
\caption{Calculations for the lowest energy states for $  J=0  $ and $  
J=2  $ in $^{38}$Si
relative to $^{34}$Si using neutron-neutron matrix
elements from SDPF-U and the renormalization procedure for both 
$^{34}$Si and $^{40}$Ca
as target nuclei, using the SHF and HO bases.
Crosses are used for calculations with $^{40}$Ca as the target nucleus. 
}
\label{(6)}
\end{figure}

Fig.\ 5 shows the same states in $^{36}$Si relative to $^{34}$Si, but 
now the comparison
includes the SDPF-U calculations and experimental data.  The CP basis 
results are not
included since they are so similar to the HO basis calculations.  We 
see that
the level scheme for $^{36}$Si is more spread out for the $  Z \geq 15  
$ SDPF-U calculation than for the
$  Z \leq 14  $ calculation, in agreement with our results discussed 
above.
Our calculations for each method are in reasonable agreement with the 
comparable
SDPF-U calculation, except for the 0$^{ + }$ state which differs by 
over 300
keV in both instances.  The experimental binding energy relative to 
$^{34}$Si is taken from a new mass measurement of
$^{36}$Si which is 140 keV higher in energy than previously measured 
\cite{mittig}.\  The excitation energies of
the $  Z \leq 14  $ SDPF-U calculation are comparable to experiment, as 
expected from an interaction fit
specifically to neutron-rich silicon isotopes.  While no
interaction reproduces the experimental data very well, general trends 
can be seen.  The calculations with
$^{40}$Ca as the target nucleus depicted by crosses result in level 
schemes that are too spread out in
comparison to the experimental data.  The reduction in the strength of 
the interaction for $^{34}$Si
using the SHF basis results in a reduction of the energy of the states 
in $^{36}$Si, especially for the
ground state (the pairing matrix elements were most reduced).  The rms 
between experiment and theory
with $^{34}$Si as the target nucleus in the SHF basis is 214 keV for 
the four states shown.  One reason
for this deviation is the lack of three body forces in the procedure.  
The inclusion of the $  NNN  $
interaction, at least via the effective two-body part, is important for 
a higher level of accuracy.
Additionally, the chosen SPEs may contribute to the deviation, which 
would be
better constrained if all the single particle states in $^{35}$Si were 
known experimentally.  For exotic nuclei,
the calculated single particle state is often above the neutron 
separation energy and determination of the
experimental single particle states may not be possible with current 
facilities.  Thus it is essential to
improve energy density functionals such that they provide reliable 
single particle energies.

The importance of using a realistic basis and an appropriate target 
nucleus to renormalize an interaction
into the nuclear medium for calculations of exotic nuclei is evident 
from the calculations shown here.
Otherwise, the interaction will be too strong and will produce 
significant overbinding even for the
two particle case, an effect that gets magnified as more particles are 
added, as seen in Fig.\ 6 for $^{38}$Si.
Only the 0$^{ + }$ and 2$^{ + }$ states are shown since the 4$^{ + }$ 
and 6$^{ + }$ states are not known experimentally, but the
binding energy is only approximately reproduced for the calculations 
with $^{34}$Si in the SHF basis.  As noted in
the $^{36}$Si case, the excitation energy of the 2$^{ + }$ state is too 
high in the SHF basis but is reproduced well by the
$  Z \leq 14  $ SDPF-U calculation for $^{38}$Si.

\section{Summary and Conclusions}

The microscopic nucleon-nucleon interaction N$^{3}$LO was renormalized 
using $  v_{low k}  $ and
many-body perturbation theory in order to
produce an effective interaction in the nuclear medium that could be 
used in a shell model code.
The renormalization was performed in three different bases: harmonic 
oscillator, core polarization,
and Skyrme Hartree-Fock.  The choice of basis can significantly affect 
the value of matrix elements, as
shown in the comparisons of $  pf  $ neutron-neutron matrix elements 
for the stable $^{40}$Ca and the
neutron-rich $^{34}$Si nuclei.  The difference results from valence 
orbits being bound by only a
few hundred keV, resulting in a long tail in the radial wavefunction 
relative to the
harmonic oscillator wavefunction.  The loosely bound orbits cause a 
reduction in the
overall strength of the interaction, an effect that becomes magnified 
as full scale shell model calculations
are performed.  Accounting for the properties of the orbits by using a 
realistic basis is essential
for an accurate description of the nuclear interaction in exotic nuclei 
as determined by the
renormalization of an $  NN  $ interaction.

\vspace{ 12pt}
\noindent 
  {\bf Acknowledgments}
Support for this work was provided from
National Science Foundation Grant PHY-0758099, from the Department of 
Energy
Stewardship Science Graduate Fellowship program through grant number
DE-FC52-08NA28752, and from the DOE UNEDF-SciDAC grant 
DE-FC02-09ER41585.

\end{document}